\documentstyle[sprocl,psfig]{article}

\newcommand{\la}{\langle}
\newcommand{\ra}{\rangle}

\newcommand{\AmS}{{\protect\the\textfont2
  A\kern-.1667em\lower.5ex\hbox{M}\kern-.125emS}}
\newcommand{\be}{\begin{equation}}
\newcommand{\ee}{\end{equation}}
\newcommand{\ben}{\begin{eqnarray}}
\newcommand{\een}{\end{eqnarray}}
\newcommand{\nn}{\nonumber}

\newcommand{\slas}[2]{{{#1}\hspace{-5pt}{/}}_{#2}}
\newcommand{\slal}[2]{{{#1}\hspace{-5pt}{/}}_{#2}}
\def\simgt{\rlap{\lower 3.5 pt\hbox{$\mathchar \sim$}}\raise 1pt \hbox {$>$}}
\def\simlt{\rlap{\lower 3.5 pt\hbox{$\mathchar \sim$}}\raise 1pt \hbox {$<$}}

\def\be{\begin{equation}}
\def\ee{\end{equation}}
\def\bea{\begin{eqnarray}}
\def\eea{\end{eqnarray}}
\begin{document}

%\vspace{-2.5cm}
%\begin{flushright}
%{\normalsize KEK-CP-107}\\
%\vspace{10mm}
%\end{flushright}

%To Prof Nick Karayiannis -- do read this:-
%If needed the word of Chapter~1, you can type in at the 
%\title{}. The words will be in caps and lowercase. 
%For chapter title can be in all caps or in caps and lowercase.
%It is up to the author to type for the case sensitive but 
%all articles must be in the same style. 
%But mostly for Review Volume are without this Chapter~1.
%Thank you
%Jessie   13/4/2000

\title{Nucleon Decay Matrix Elements from Lattice QCD}

\author{Yoshinobu~Kuramashi for 
JLQCD Collaboration\footnote{JLQCD Collaboration:
S.~Aoki, M.~Fukugita, S.~Hashimoto, K.-I.~Ishikawa, N.~Ishizuka,
Y.~Iwasaki, K.~Kanaya, T.~Kaneda, S.~Kaya, Y.~K., M.~Okawa,
T.~Onogi, S.~Tominaga, N.~Tsutsui, A.~Ukawa, N.~Yamada, T.~Yoshi\'e.}}

\address{Institute of Particle and Nuclear Studies,\\ 
         High Energy Accelerator Research Organization(KEK),\\ 
         Tsukuba, Ibaraki 305-0801, Japan\\
         E-mail: yoshinobu.kuramashi@kek.jp} 

%\author{A. N. OTHER}
%
%\address{Department of Physics, Theoretical Physics, 1 Keble Road,\\
%Oxford OX1 3NP, England\\E-mail: other@tp.ox.uk}


\maketitle\abstracts{ 
We present a GUT-model-independent calculation of hadron matrix elements
for all dimension-six operators associated with baryon number 
violating processes using lattice QCD. 
%The calculation is performed with the
%Wilson quark action in the quenched approximation at 
%$\beta=6/g^2=6.0$ on a $28^2\times 48\times 80$ lattice.
Our results cover all the matrix elements required to estimate
the partial lifetimes of (proton,neutron)$\rightarrow$($\pi,K,\eta$)
+(${\bar \nu},e^+,\mu^+$) decay modes.
We point out the necessity of disentangling two form factors that contribute
to the matrix elements; previous calculations did not make the separation, 
which led to an underestimate of the physical matrix elements.
With a correct separation, we find that the matrix elements
have values $3-5$ times larger than the smallest
estimates employed in phenomenological analyses of the nucleon
decays, which gives stronger constraints on
GUT models.
We also find that the values of the matrix elements are comparable
with the tree-level predictions of chiral Lagrangian. 
}

\section{Introduction}
%\subsection{Producing the Hard Copy}\label{subsec:prod}

Nucleon decay is one of the most exciting predictions of
grand unified theories (GUTs) regardless of the existence
of supersymmetry (SUSY). 
Although none of the decay modes have been detected up to now,
experimental efforts over the years have pushed 
the lower limit on the partial lifetimes of the nucleon\cite{pd_ex}, 
which can give
a constraint on (SUSY-)GUTs. 
Moreover, exciting plans for the next generation 
of the Super-Kamiokande experiment
to search the nucleon decay events are proposed 
in this conference\cite{sk_next}. 

On the theoretical side predictions of the nucleon partial lifetimes
suffer from various uncertainties. In general the partial lifetime 
of the nucleon decay process $N\rightarrow PS+{\bar l}$, where
$N$, $PS$ and ${\bar l}$ denote nucleon, pseudoscalar
meson and antilepton respectively, is
given by
\be
\tau\propto 
|\la PS | {\cal O}^{\slal{B}{}} |N\ra|^{-2} \cdot |F_{\rm GUT}|^{-2},
\label{eq:lifetime}
\ee
with ${\cal O}^{\slal{B}{}}$
the baryon number violating operator that appears in the 
low-energy effective Lagrangian of (SUSY-)GUTs. $F_{\rm GUT}$ denotes
some function in terms of parameters defined in the (SUSY-)GUT models.
Although the precise numbers of the nucleon partial lifetimes
depend on the details of the
theories beyond the standard model, which contain many unknown parameters,
another main source of uncertainty is found in the evaluation
of the hadronic part
$\la PS | {\cal O}^{\slal{B}{}} |N\ra$ in eq.~(\ref{eq:lifetime}). 
The matrix elements have been estimated by
employing various QCD models. Their results, however,
%scatter over a factor of ten\cite{model_wf}.
%Our aim is a precise determination of the nucleon decay matrix elements
%from the first principles using lattice QCD.
scatter over the range whose minimum and maximum values differ by 
a factor of ten\cite{model_wf}.
Therefore, a precise determination of the nucleon decay matrix elements
from the first principles using lattice QCD is of extreme importance.

Most important feature of our calculation is GUT 
model independence.
%Although the precise numbers of the nucleon partial lifetimes
%depend on the details of the
%theories beyond the standard model, a GUT-model-independent 
%analysis can be applied to the hadronic part 
%$\la PS | {\cal O}^{\slal{B}{}} |N\ra$
%in eq.~(\ref{eq:lifetime}):   
All dimension-six operators associated with baryon number 
violating processes 
are classified into four types under the requirement of
SU(3)$\times$SU(2)$\times$U(1) invariance 
at low-energy scales\cite{op_6,op_4}. 
If one specifies the decay processes of interest,
namely the processes among (proton,neutron)$\rightarrow$($\pi,K,\eta$)
+(${\bar \nu},e^+,\mu^+$), we can list a complete set of
independent matrix elements in QCD, and we calculate 
all the matrix elements.

We employed two methods to evaluate the nucleon decay matrix elements 
in lattice QCD. One is the indirect method in which 
the matrix element 
$\la PS | {\cal O}^{\slal{B}{}} |N \ra$ is estimated 
from $\la 0 | {\cal O}^{\slal{B}{}} |N\ra$ 
measured on the lattice employing the 
tree-level results of chiral Lagrangian.
The other is the direct method which directly measures 
$\la PS | {\cal O}^{\slal{B}{}} |N\ra$ on the lattice.
In the previous lattice QCD studies the indirect method\cite{hara,bowler}
gave values of the matrix elements 2 or 3 times smaller than those obtained 
by the direct method\cite{gavela,jlqcd_98}.
Recently, however, we pointed out\cite{jlqcd_00} 
that the nucleon decay matrix elements 
$\la PS | {\cal O}^{\slal{B}{}} |N\ra$ allow the contributions of 
two form factors for general lepton momentum, which should be disentangled 
in the direct method, and
this explains the discrepancy between the direct and indirect 
estimations of the matrix elements 
found in the previous studies\cite{gavela,jlqcd_98} 
where the separation was not made. 

In this report we first  
formulate the method to calculate  
the nucleon decay matrix elements using the lattice QCD 
in Sec.~\ref{sec:formulation}.
Our results and conclusions are given in Secs.~\ref{sec:results} 
and \ref{sec:conclusions}.
More details are found in Ref.~10.
%~\protect{\cite{jlqcd_00}}.

\section{Formulation of the method}
\label{sec:formulation}

\subsection{Independent matrix elements for nucleon decays}
%\label{subsec:prod}

The low energy effective theory in  
the baryon number violating processes is described
in terms of  SU(3)$\times$SU(2)$\times$U(1) gauge symmetry
based on the strong and the electroweak interactions,
which enables us to make a GUT-model-independent analysis.
Our interest is focused on the dimension-six operators
which are the lowest dimensional ones 
in the low energy effective Hamiltonian:
operators must contain at least three quark fields to form SU(3) color
singlet, and then an additional lepton field is
required to construct a Lorentz scalar.
Higher-dimensional operators 
are suppressed by inverse powers of heavy particle mass
characterized by the theory beyond the standard model.

All dimension-six operators are classified
into the four types under the
requirement of SU(3)$\times$SU(2)$\times$U(1) invariance 
\cite{op_6,op_4}:
\ben
{\cal O}^{(1)}_{abcd}&=&
({\bar D}^c_{i aR}U_{j bR})
({\bar Q}^c_{\alpha k cL}L_{\beta dL})
\epsilon_{ijk}\epsilon_{\alpha\beta},  
\label{eq:op_1}\\
{\cal O}^{(2)}_{abcd}&=&
({\bar Q}^c_{\alpha i aL}Q_{\beta j bL})
({\bar U}^c_{k cR}L_{dR})
\epsilon_{ijk}\epsilon_{\alpha\beta},  
\label{eq:op_2}\\
{\cal O}^{(3)}_{abcd}&=&
({\bar Q}^c_{\alpha i aL}Q_{\beta j bL})
({\bar Q}^c_{\gamma k cL}L_{\delta dL})
\epsilon_{ijk}\epsilon_{\alpha\delta}\epsilon_{\beta\gamma}, 
\label{eq:op_3}\\ 
{\cal O}^{(4)}_{abcd}&=&
({\bar D}^c_{i aR}U_{j bR})
({\bar U}^c_{k cR}L_{dR})
\epsilon_{ijk},
\label{eq:op_4}
\een 
where ${\bar \psi}^c=\psi^T C$ 
with $C=\gamma_4\gamma_2$ the charge conjugation matrix;
$i$, $j$ and $k$ are SU(3) color indices; 
$\alpha$, $\beta$, $\gamma$ and $\delta$ are SU(2) indices; 
$a$, $b$, $c$ and 
$d$ are generation indices; $L_L$ and $Q_L$ are generic lepton and quark
SU(2) doublets with the left-handed projection $P_L=(1-\gamma_5)/2$;
$L_R$, $U_R$, and $D_R$ are generic charged lepton and quark SU(2)
singlets with the right-handed projection $P_R=(1+\gamma_5)/2$.
Fierz transformations are used to eliminate all the vector
and tensor Dirac structures in eqs.~(\ref{eq:op_1})$-$(\ref{eq:op_4}).

Our interest exists in the decay processes from the nucleon to one
pseudoscalar meson: (proton,neutron)$\rightarrow$($\pi,K,\eta$)
+(${\bar \nu},e^+,\mu^+$). Once these decay modes are specified,
we can list the set of independent matrix elements
in QCD from the operators of eqs.~(\ref{eq:op_1})$-$(\ref{eq:op_4}):
\ben
&&\langle \pi^0|\epsilon_{ijk}
({u^i}^T CP_{R,L}d^j) P_L u^k|p\rangle, 
\label{eq:indme_1}\\ 
&&\langle \pi^+|\epsilon_{ijk}
({u^i}^T CP_{R,L}d^j) P_L d^k|p\rangle, 
\label{eq:indme_2}\\
&&\langle K^0|\epsilon_{ijk}
({u^i}^T CP_{R,L}s^j) P_L u^k|p\rangle, 
\label{eq:indme_3}\\
&&\langle K^+|\epsilon_{ijk}
({u^i}^T CP_{R,L}s^j) P_L d^k|p\rangle, 
\label{eq:indme_4}\\
&&\langle K^+|\epsilon_{ijk}
({u^i}^T CP_{R,L}d^j) P_L s^k|p\rangle, 
\label{eq:indme_5}\\
&&\langle K^0|\epsilon_{ijk}
({u^i}^T CP_{R,L}s^j) P_L d^k|n\rangle, 
\label{eq:indme_6}\\
&&\langle \eta |\epsilon_{ijk}
({u^i}^T CP_{R,L}d^j) P_L u^k|p\rangle,
\label{eq:indme_7} 
\een
where we assume SU(2) isospin symmetry $m_u=m_d$ and use
the relations
\ben
\la PS | {\cal O}_{LR} |N \ra&=&\la PS | {\cal O}_{RL} |N \ra,
\label{eq:lr=rl}\\
\la PS | {\cal O}_{RR} |N \ra&=&\la PS | {\cal O}_{LL} |N \ra,
\label{eq:rr=ll}
\een
due to the parity invariance.
All we have to calculate in lattice QCD are these 14 matrix elements.
Other matrix elements can be obtained by using
the exchange of the up and down quarks and the relations of 
eqs.~(\ref{eq:lr=rl}) and (\ref{eq:rr=ll}). 

\nopagebreak

\subsection{Form factors in nucleon decay matrix elements}
\label{subsec:ff}

Under the requirement of Lorentz and parity invariance,
the matrix elements 
between the nucleon and
the pseudoscalar meson in eqs.~(\ref{eq:indme_1})$-$(\ref{eq:indme_7}) 
can have two form factors:
\be
\langle PS(\vec{p})|{\cal O}_L^{\slal{B}{}}|N^{(s)}(\vec{k})\rangle
=P_L\left(W_0(q^2)-W_q(q^2) i{\slas{q}{}}\right) u^{(s)},
\label{eq:ff}
\ee
where ${\cal O}^{\slal{B}{}}_L$ represents the three-quark operator
projected to the left-handed chiral state, 
$u^{(s)}$ denotes the Dirac spinor for nucleon  
with either the up ($s=1$) or down ($s=2$) spin state,
and $q^2$ is the momentum squared of the out-going antilepton.
The contribution of the $W_q$ term in  eq.(\ref{eq:ff}) 
is negligible in the physical decay amplitude, because its contribution 
is of the order of the lepton mass $m_l$ 
after the multiplication with antilepton spinor. 
However, since the relative magnitude of the two form factors $W_0$ and 
$W_q$ is {\it a priori} not known, we have to disentangle these 
two form factors in the lattice QCD calculation. 
%in which the antilepton 
%momentum is $O(1)$ in lattice units. 
Hereafter we refer to $W_0$ and $W_q$ as relevant 
and irrelevant form factor respectively.

In our lattice QCD calculation we choose $\vec{k}=\vec{0}$ for 
the nucleon spatial momentum and $\vec{p}=\vec{k}-\vec{q}\neq \vec{0}$ for
the PS meson.
In this case the Dirac structure of the right hand side 
in eq.~(\ref{eq:ff}) is given by
\ben
&&\left( W_0 - W_q i{\slas{q}{}}\right) u^{(s)}\nn\\&=&
\left(
\begin{array}{cc}
     W_0 - iq_4 W_q       
&   -W_q {\vec q}\cdot\vec{\sigma}    \\
     W_q {\vec q}\cdot\vec{\sigma}  
&    W_0 + i q_4 W_q
\end{array}
\right) u^{(s)}
\label{eq:dstructure}\\
&=&
\left(
\begin{array}{cc}
    W_0 + (m_N-\sqrt{m_{PS}^2+{\vec p}^2}) W_q  
&   W_q {\vec p}\cdot\vec{\sigma}   \\
   -W_q {\vec p}\cdot\vec{\sigma}  
&   W_0 - (m_N-\sqrt{m_{PS}^2+{\vec p}^2}) W_q  
\end{array}
\right) u^{(s)},\nn
\een
where $W_0-W_qi{\slas{q}{}}$ is expressed by a $2\times 2$ block notation;
$\vec{\sigma}$ are the Pauli matrices, and 
${u^{(s)}}^T=(1,0,0,0)$ or $(0,1,0,0)$. 
Using the $(1,1)$ and $(2,1)$ components in the $2\times 2$ block notation
of eq.~(\ref{eq:dstructure}), where the other components vanish, 
we can extract the relevant form factor $W_0$.

%It is important to observe that the upper components of 
%$( W_0 - W_q i{\slas{q}{}}) u^{(s)}$ 
%are linear combinations of the 
%relevant and irrelevant form factors, 
%while the lower components contain only the irrelevant one.
%Therefore, we can extract the relevant form factor $W_0$ from the
%upper components by subtracting the contribution of 
%the irrelevant form factor
%$W_q$ with the use of the lower components.

The need for the separation of the contribution of the irrelevant 
form factor was not recognized in the previous studies with the 
direct method\cite{gavela,jlqcd_98}.  The values found in these 
studies correspond to  $W_0-iq_4W_q$ instead of $W_0$.  We examine how 
much this affects the estimate of the matrix elements 
in Sec.~\ref{sec:results}.

%Let us add several technical comments: 
%(i) The separation procedure described above cannot be applied to
%the case of ${\vec p}={\vec k}={\vec 0}$ because of 
%vanishing lower components.
%(ii) Another possible choice of momenta for disentangling the relevant 
%and irrelevant form factors is given by ${\vec k}\ne{\vec 0}$ and 
%${\vec p}={\vec 0}$.  In this case, however, we cannot achieve
%$-q^2=m_l^2$.

\subsection{Calculational methods}
\label{subsec:calmethod}

The nucleon decay matrix elements 
of eq.~(\ref{eq:ff}) are calculated with two methods referred
to as the direct and the indirect ones.  
The former is to extract the matrix elements 
from the three-point
function of the nucleon, the PS meson and the baryon number violating
operator.  
The latter is to estimate them with the aid of chiral Lagrangian,
where we have two unknown parameters to be determined
by the lattice QCD calculation. 
 
In the direct method
we calculate the following ratio of the hadron three-point function
divided by the propagators of the pseudoscalar meson and the nucleon:
\ben
&&R(t,t^\prime)
\nn\\&=&
\frac{\sum_{{\vec x},{\vec x}^\prime} 
      {\rm e}^{i{\vec p}\cdot({\vec x}^\prime-{\vec  x})}
      \la { J}_{PS}({\vec x}^\prime,t^\prime) 
          {\hat {\cal O}}^{\slal{B}{}}_{L,\gamma}({\vec x},t) 
          \bar{ J}_{N,s}(0) \ra}
     {\sum_{{\vec x},{\vec x}^\prime}
      {\rm e}^{i{\vec p}\cdot({\vec x}^\prime-{\vec x})}
      \la { J}_{PS}({\vec x}^\prime,t^\prime) 
          { J}_{PS}^{\dagger}({\vec x},t) \ra
      \sum_{\vec x}
      \la { J}_{N,s}({\vec x},t) 
      \bar{ J}_{N,s}(0) \ra }
\sqrt{Z_{PS}} \sqrt{Z_N} \nn \\
&\longrightarrow& 
\frac{1}{L_x L_y L_z}
\la PS(\vec{p})|{\hat {\cal O}}^{\slal{B}{}}_{L,\gamma}
|N^{(s)}({\vec k}={\vec 0})\ra
\;\;\;\;\;\;\;\; t^\prime \gg t \gg 0.
\label{eq:ratio}
\een
Here ${\hat {\cal O}}^{\slal{B}{}}_{L,\gamma}$ denotes
the renormalized operator in the 
naive dimensional regularization(NDR) with the ${\overline{\rm MS}}$ 
subtraction scheme, and ${ J}_{PS}$ and ${ J}_{N,s}$ are interpolating
fields for the PS meson and the nucleon, respectively.
The subscripts $\gamma$ and $s$ are spinor indices; we can specify the
spin state of the initial nucleon at rest by choosing
$s=1$ or $2$. 
$Z_{PS}$ and $Z_N$ represent the residues of the PS meson propagator
and the nucleon propagator.
$L_xL_yL_z$ is the spatial volume of the lattice.
In lattice QCD calculation we evaluate the Green functions numerically 
by the Monte Carlo method with supercomputer.    
We move the baryon number violating operator 
${\hat {\cal O}}^{\slal{B}{}}_{L,\gamma}$ 
in terms of $t$ between the nucleon source placed at $t=0$ and 
the PS meson sink fixed at some $t^\prime$ well separated from 
$t=0$. Under the condition that the baryon number violating operator
is sufficiently distanced from both the PS meson field and the nucleon field, 
which is required to avoid excited state contaminations,
we can extract the matrix elements normalized 
by the spatial volume of the lattice.

The indirect method uses the tree-level results of chiral Lagrangian,
which are given by
\ben
\langle \pi^0|(ud_R) u_L|p\rangle
&=&\alpha P_L u_p\left[1+D+F\right]/(\sqrt{2}f),
\label{eq:chpt_1_rl_q}\\ 
\langle \pi^0|(ud_L) u_L|p\rangle
&=&\beta P_L u_p\left[1+D+F\right]/(\sqrt{2}f),
\label{eq:chpt_1_ll_q}\\ 
\langle \pi^+|(ud_R) d_L|p\rangle
&=&\alpha P_L u_p\left[1+D+F\right]/f,
\label{eq:chpt_2_rl_q}\\
\langle \pi^+|(ud_L) d_L|p\rangle
&=&\beta P_L u_p\left[1+D+F\right]/f,
\label{eq:chpt_2_ll_q}\\
\langle K^0|(us_R) u_L|p\rangle 
&=&\alpha P_L u_p\left[-1-(D-F)m_{N/B}\right]/f,
\label{eq:chpt_3_rl_q}\\
\langle K^0|(us_L) u_L|p\rangle 
&=&\beta P_L u_p\left[1-(D-F)m_{N/B}\right]/f,
\label{eq:chpt_3_ll_q}\\
\langle K^+|(us_R) d_L|p\rangle
&=&\alpha P_L u_p\left[2Dm_{N/B}\right]/(3f),
\label{eq:chpt_4_rl_q}\\
\langle K^+|(us_L) d_L|p\rangle
&=&\beta P_L u_p\left[2Dm_{N/B}\right]/(3f),
\label{eq:chpt_4_ll_q}\\
\langle K^+|(ud_R) s_L|p\rangle 
&=&\alpha P_L u_p\left[3+(D+3F)m_{N/B}\right]/(3f),
\label{eq:chpt_5_rl_q}\\
\langle K^+|(ud_L) s_L|p\rangle 
&=&\beta P_L u_p\left[3+(D+3F)m_{N/B}\right]/(3f),
\label{eq:chpt_5_ll_q}\\
\langle K^0|(us_R) d_L|n\rangle
&=&\alpha P_L u_n\left[-3-(D-3F)m_{N/B}\right]/(3f),
\label{eq:chpt_6_rl_q}\\
\langle K^0|(us_L) d_L|n\rangle
&=&\beta P_L u_n\left[3-(D-3F)m_{N/B}\right]/(3f),
\label{eq:chpt_6_ll_q}\\
\langle \eta |(ud_R) u_L|p\rangle
&=&\alpha P_L u_p\left[-1-(D-3F)\right]/(\sqrt{6}f),
\label{eq:chpt_7_rl_q}\\ 
\langle \eta |(ud_L) u_L|p\rangle
&=&\beta P_L u_p\left[3-(D-3F)\right]/(\sqrt{6}f),
\label{eq:chpt_7_ll_q}
\een
where $f$ is the pion decay constant; $F$ and $D$ parameters 
are determined from experimental results
of the semileptonic baryon decays;
$m_{N/B}=m_N/m_B$ with $m_B\equiv m_\Sigma\simeq m_\Lambda$.
We use $\langle PS |(\psi_1 {\psi_2}_{R,L}) {\psi_3}_L|N\rangle$ 
as a shortened form of $\langle PS|\epsilon_{ijk}
({\psi_1^i}^T CP_{R,L}{\psi_2^j}) 
P_L{\psi_3^k}|N\rangle$.

The expressions in eqs.~(\ref{eq:chpt_1_rl_q})$-$(\ref{eq:chpt_7_ll_q})
contain  two unknown coefficients $\alpha$ and $\beta$, which are
to be determined by lattice QCD calculation.
The definitions of $\alpha$ and$\beta$ parameters are given by 
\ben
\la 0 | \epsilon_{ijk}({u^i}^T C P_R d^j) P_L u^k 
| p^{(s)} \ra &=& \alpha P_L u^{(s)}, 
\label{eq:alpha}\\
\la 0 | \epsilon_{ijk}({u^i}^T C P_L d^j) P_L u^k 
| p^{(s)} \ra &=& \beta  P_L u^{(s)},
\label{eq:beta}
\een
where operators are renormalized in the NDR scheme.
$p$ denotes the proton state.
These matrix elements are obtained from the ratio of two-point
functions:
\ben
R^{\alpha\beta}(t)&=&
\frac{\sum_{\vec x}
      \la \epsilon_{ijk}({u^i}^T C P_{R,L} d^j) P_L u^k({\vec x},t)
      \bar{ J}_{p,s}(0)\ra}
     {\sum_{\vec x}
      \la { J}_{p,s}({\vec x},t)
      \bar{ J}_{p,s}(0)\ra}
      \sqrt{Z_N} \nn\\
      &\longrightarrow&
      \la 0 | \epsilon_{ijk}(u^i C P_{R,L} d^j) P_L u^k | p^{(s)} \ra
\;\;\;\;\;\;\;\; t\gg 0 . 
\label{eq:ratio_ab}
\een
Incorporating the $\alpha$ and $\beta$ values
determined by the lattice calculation  into the
tree-level results of chiral Lagrangian, 
we can obtain the values of the
nucleon decay matrix elements in eqs.~(\ref{eq:indme_1})$-$(\ref{eq:indme_7}).

\section{Results} 
\label{sec:results}

\subsection{Details of numerical simulation}

Our calculation is carried out with the Wilson quark action in quenched QCD 
at $\beta=6.0$ on a $28^2\times 48\times 80$ lattice. 
Gauge configurations are generated with the single plaquette
action separated by 2000 pseudo heat-bath sweeps. 
We analyzed 100 configurations for the calculation of the 
nucleon decay matrix elements after the thermalization of 22000 sweeps.
The four hopping parameters $K=0.15620$, $0.15568$, 
$0.15516$ and $0.15464$ 
are adopted such that the physical point for the $K$ meson can be
interpolated.
The critical hopping parameter $K_c=0.15714(1)$ is determined by
extrapolating the results of $m_\pi^2$ at the four hopping parameters 
linearly in $1/2K$ to $m_\pi^2=0$. The $\rho$ meson mass at the 
chiral limit is used to determine the inverse lattice spacing 
$a^{-1}=2.30(4)$GeV with $m_\rho=770$MeV as input.
The strange quark mass $m_s a=0.0464(16)$($K_s=0.15488(7)$), 
which is estimated from the
experimental ratio $m_K/m_\rho=0.644$, is in the middle of $K=0.15516$ and
$K=0.15464$.

We calculate the ratio of eq.~(\ref{eq:ratio}) with 
the nucleon field fixed at $t=0$ and the PS meson field
at $t=29$.
Four spatial momenta ${\vec p}a=(0,0,0)$, $(\pi/14,0,0)$, $(0,\pi/14,0)$
and $(0,0,\pi/24)$ are imposed on the PS meson in the final
state. For the ${\vec p}\ne {\vec 0}$ cases
we provide different quark masses for the valence quark connecting 
$J_{PS}$ and ${\hat {\cal O}}^{\slal{B}{}}_{L,\gamma}$ 
and the other valence quarks:  
$m_2$ is for the former and $m_1$ for the latter.
By this assignment
we can distinguish the strange quark mass from the
up and down quark mass. 
%As explained in Sec.~\ref{subsec:ff}, we cannot disentangle
%the relevant form factor from the irrelevant one
%in the case of the PS meson at rest, where we take
%only the degenerate quark mass $m_1=m_2$.

\begin{figure}[t]
\begin{minipage}[t]{55mm}
\centering{
\hskip -0.0cm
\psfig{file=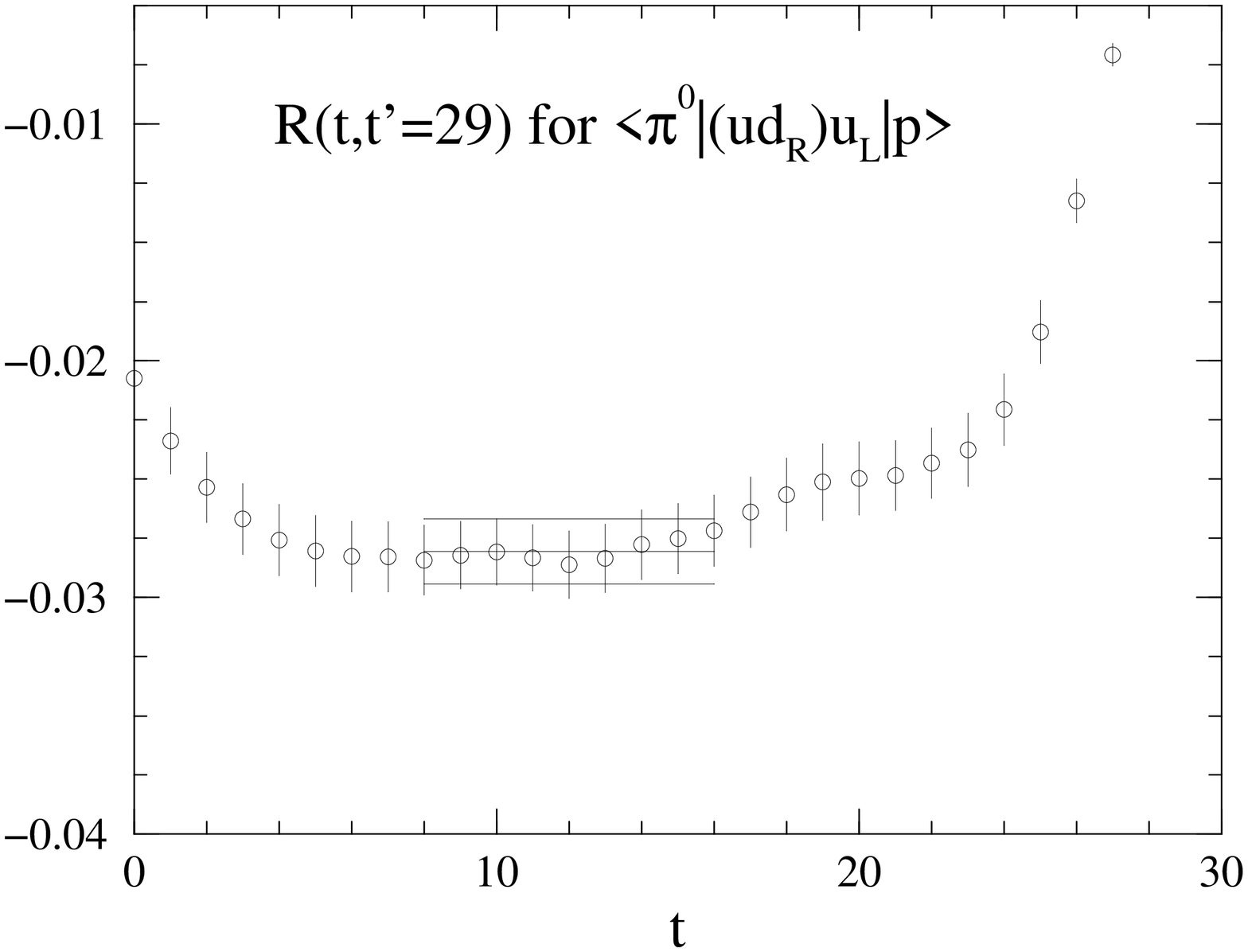,width=53mm,angle=-90}
}
%\vskip -10mm}
\caption{Ratio $R(t,t^\prime=29)$ for the relevant form factor $W_0$ in
$\la \pi^0|(ud_R) u_L|p\ra$. 
Solid lines denote the fitting results with an error band of  one standard
deviation.} 
\label{fig:ratio_3pt_xy}
\end{minipage}
\hspace{\fill}
\begin{minipage}[t]{57mm}
\centering{
\hskip -0.0cm
\psfig{file=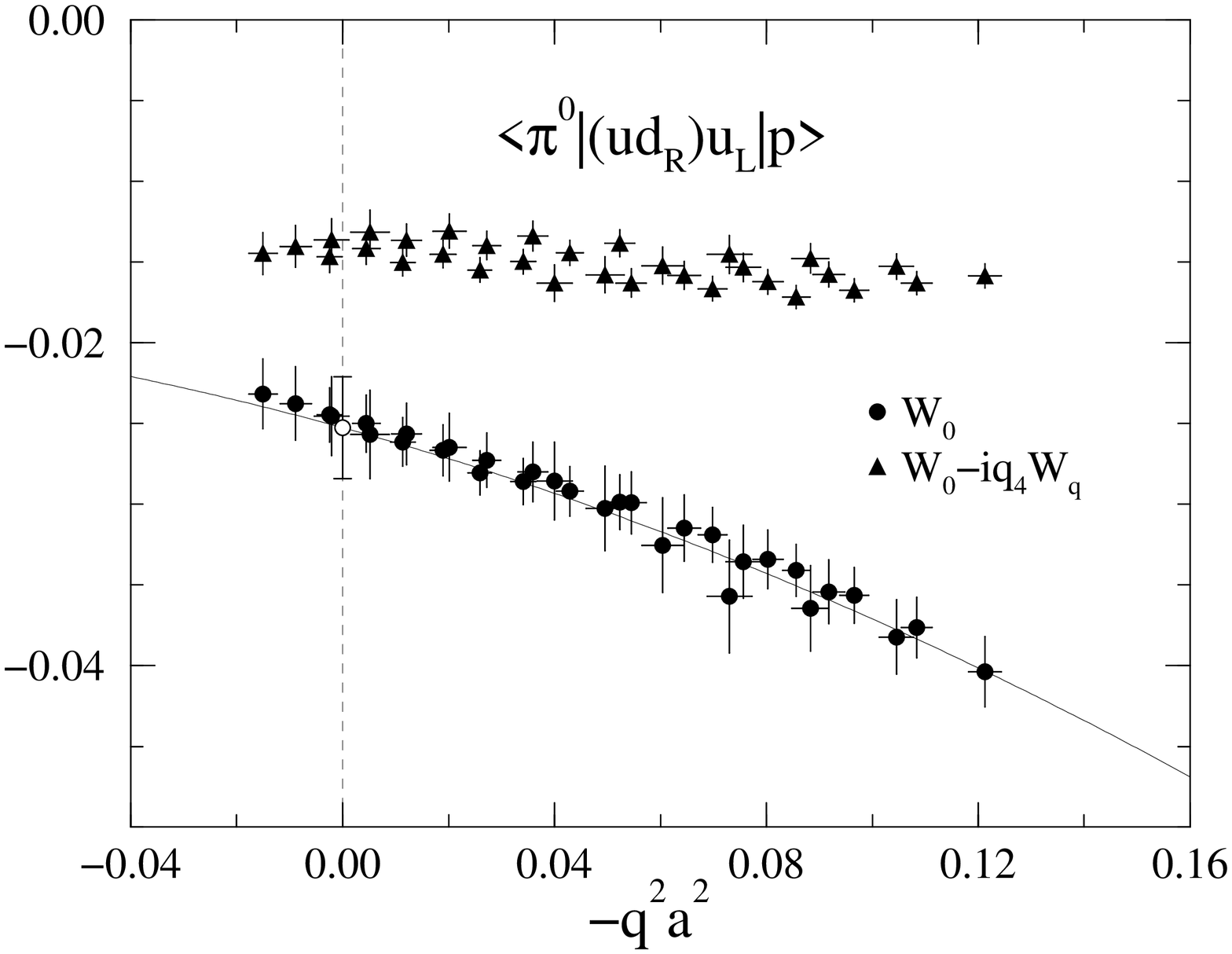,width=55mm,angle=-90}
}
%\vskip -10mm  }
\caption{$-q^2 a^2$ dependences for $W_0$
in $\la \pi^0|(ud_R) u_L|p\ra$.
Combination of form factors $W_0-iq_4 W_q$ 
is also plotted for comparison. 
Solid line denotes the function 
$c_0+c_1\cdot (-q^2 a^2)+c_2\cdot (-q^2a^2)^2$.} 
\label{fig:qfit1}
\end{minipage}
\vspace{-5mm}
\end{figure}

\subsection{Nucleon decay matrix elements 
with direct and indirect methods}

We first present the results of the nucleon decay matrix elements
obtained by the direct method. 
In Fig.~\ref{fig:ratio_3pt_xy} we show time
dependence of $R(t,t^\prime=29)$ with $|{\vec p}|a=\pi/14$ for
the matrix element $\la \pi^0|\epsilon_{ijk}
({u^i}^T CP_{R}d^j) P_L u^k|p\ra$ 
as a representative case. The result of constant fit
is depicted by the set of three  horizontal lines.
We choose the fitting range to be $8 \le t \le 16$
for all the matrix elements of 
eqs.~(\ref{eq:indme_1})$-$(\ref{eq:indme_7}) such that
the excited state contaminations in the nucleon
and PS meson states can be avoided simultaneously.

Figure~\ref{fig:qfit1} shows $-q^2a^2$ dependence
of the relevant form factor $W_0(q^2)$ in the matrix element 
$\la \pi^0|\epsilon_{ijk}({u^i}^T CP_{R}d^j) P_L u^k|p\ra$, where 
the operators are renormalized with the NDR scheme at $\mu=1/a$.
The fitting result in Fig.~\ref{fig:ratio_3pt_xy} corresponds to the data at
$-q^2a^2=0.0259(22)$ in Fig.~\ref{fig:qfit1}.
The combination $W_0-iq_4 W_q$ is also plotted in Fig.~\ref{fig:qfit1} 
for comparison, which
is obtained by following the method in Ref.~8.
%~\cite{gavela}.
The magnitude of  $W_0(q^2)$ is more than two times larger than that
of $W_0(q^2)-iq_4 W_q(q^2)$.

To interpolate the relevant form factor $W_0$ to $q^2=0$, 
which is the on-shell point of outgoing antilepton,
we employ the following fitting function:
\be
c_0+c_1\cdot(-q^2) +c_2\cdot (-q^2)^2 +c_3\cdot m_1 +c_4\cdot m_2,
\label{eq:qfit}
\ee
where we assume that the form factor could have the $m_1$ and $m_2$
dependences through the nucleon and PS meson masses. 
We extrapolate  $m_1$ and $m_2$ to the chiral limit
for the matrix elements of 
eqs.~(\ref{eq:indme_1}), (\ref{eq:indme_2}) and (\ref{eq:indme_7}), while
$m_2$ is interpolated to the physical strange quark mass
with $m_1$ taken to the chiral limit  
for the matrix elements of eqs.~(\ref{eq:indme_3})$-$(\ref{eq:indme_6}).
The solid line and the open circle at $-q^2a^2=0$ 
in Fig.~\ref{fig:qfit1} denote
the fitting result of the data employing 
the function of eq.~(\ref{eq:qfit}), where we find that
the charged lepton masses $m_e^2 a^2=4.9\times 10^{-8}$ 
and $m_\mu^2 a^2=2.1\times 10^{-3}$ 
are negligible in the current numerical statistics.
The solid line expresses the function 
$c_0+c_1\cdot(-q^2)+c_2\cdot(-q^2)^2$ 
employing the fitting results of $c_0$, $c_1$ and $c_2$.

Let us turn to the results
obtained by the indirect method, which uses the tree-level results 
of chiral Lagrangian in 
eqs.~(\ref{eq:chpt_1_rl_q})$-$(\ref{eq:chpt_7_ll_q}). 
The $\alpha$ and $\beta$ parameters at each hopping parameter 
are extracted from a constant fit of the ratio of eq.~(\ref{eq:ratio_ab}).
Applying linear fits to the data as a function of quark mass 
$m_qa=(1/K-1/K_c)/2$, we obtain
$\alpha({\rm NDR},1/a)=-0.015(1)$GeV$^3$ and 
$\beta({\rm NDR},1/a)=0.014(1)$GeV$^3$ 
in the chiral limit with the use of $a^{-1}=2.30(4)$GeV. 

\begin{figure}[t]
\centering{
\hskip -0.0cm
\psfig{file=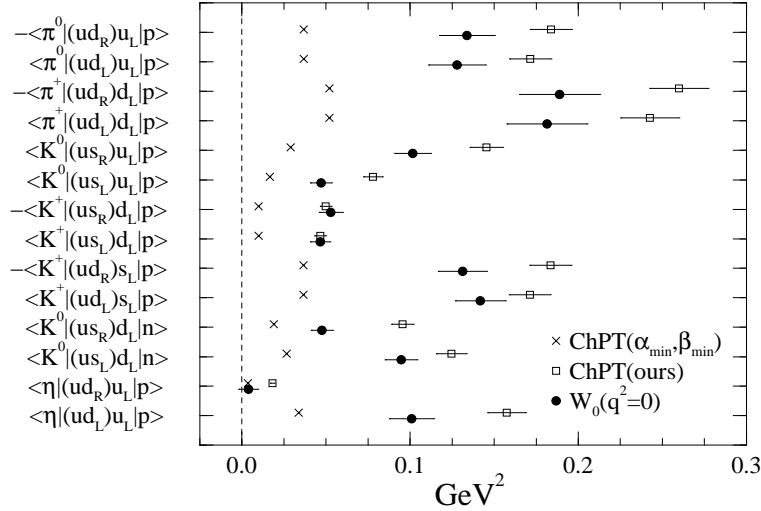,width=100mm,angle=-90}
%\vskip -10mm  
}
\caption{Comparison of relevant form factors with tree-level predictions
of ChPT. Crosses denote the ChPT results with
$|\alpha|=|\beta|=0.003$GeV$^3$. For numerical values see 
Ref.~10.
}
\label{fig:summary}
%\vspace{-8mm}
\end{figure}

In phenomenological GUT model analyses of the nucleon decays, 
the values $|\alpha|=|\beta|=0.003$GeV$^3\;\;$\cite{ab_min} 
are conservatively 
taken as these are the smallest estimate among various
QCD model calculations\cite{model_wf}.  
The previous 
lattice calculations, however, indicated values of these parameters 
considerably larger than the minimum model 
estimate above\cite{hara,bowler,gavela}. 
Our results, significantly improved over the previous ones 
due to the use of higher statistics, larger spatial size, 
lighter quark masses and smaller lattice spacing, 
have confirmed this trend:
the values we obtained are about five times larger than 
$|\alpha|=|\beta|=0.003$GeV$^3$.

In Fig.~\ref{fig:summary} we compare the nucleon decay matrix
elements obtained by the direct
method with those by the indirect one using the tree-level results
of chiral Lagrangian (squares), where we employ the expressions of
eqs.~(\ref{eq:chpt_1_rl_q})$-$(\ref{eq:chpt_7_ll_q}) 
with $\alpha({\rm NDR},1/a)=-0.015(1)$GeV$^3$,
$\beta({\rm NDR},1/a)=0.014(1)$GeV$^3$, $f_\pi=0.131$GeV, $m_N=0.94$GeV,
$m_B=1.15$GeV, $D=0.80$ and $F=0.47$\cite{fd}.
We observe that the two set of results are roughly comparable.  
This leads us to consider that the large discrepancy between the results 
of the two methods found in Refs.~8,9
%\cite{gavela,jlqcd_98} 
is mainly due to the neglect of the
$W_q(q^2)$ term in eq.~(\ref{eq:ff}).
  
It is also intriguing to compare our results with 
the tree-level predictions of chiral Lagrangian 
with $|\alpha|=|\beta|=0.003$GeV$^3$ (crosses).
Our results with the direct method are $3-5$ times larger than the
smallest estimates except $\langle \eta |(ud_R) u_L|p\rangle$.  
Hence they are expected to give stronger 
constraints on the parameters of GUT models. 

\section{Conclusions}
\label{sec:conclusions}

In this article we have reported progress in the lattice study of
the nucleon decay matrix elements.  
In order to enable a GUT-model-independent analysis of the nucleon decay, 
we have extracted the form factors of all the independent matrix elements
relevant for the (proton,neutron)$\rightarrow$($\pi,K,\eta$)
+(${\bar \nu},e^+,\mu^+$) decay processes without invoking
chiral Lagrangian. 

We have also pointed out the necessity of separating out the 
contribution of an irrelevant form factor in lattice calculations 
for a correct estimate of the matrix elements at the physical point. 
With this separation, the matrix elements obtained from the three-point
functions are roughly 
comparable with the tree-level predictions of chiral Lagrangian
with the $\alpha$ and $\beta$ parameters 
determined on the same lattice.
The magnitude of the matrix elements, however, are 
3 to 5 times larger than those with the smallest
estimate of $\alpha$ and $\beta$ among various QCD model
calculations.
Our results would stimulate phenomenological interests as
the larger values of the nucleon decay matrix elements
can give more stringent constraints on GUT models.

%The ultimate goal of lattice QCD calculations 
%of the nucleon decay matrix elements is to
%determine the matrix elements precisely
%with control over possible systematic errors.
Major systematic errors conceivably affecting our present results are 
the scaling violations and the quenching effects.
The former can be investigated by repeating the simulation
at several lattice spacings;
the latter is eliminated 
once configurations are generated with dynamical
quarks, where it is straightforward to apply our method. 
We leave these points to future studies.

\section*{Acknowledgments}
This work is supported by the Supercomputer Project No.45 (FY1999)
of High Energy Accelerator Research Organization (KEK),
and also in part by the Grants-in-Aid of the Ministry of 
Education (Nos. 09304029, 10640246, 10640248, 10740107, 10740125,
11640294, 11740162).

\end{document}